\documentclass[aps,
twocolumn,reprint,
showpacs,preprintnumbers,
amsmath,amssymb,floatfix,a4paper,longbibliography,superscriptaddress]{revtex4-2} 

\usepackage{amsmath}
\usepackage{amsfonts}
\usepackage{amssymb}
\usepackage{bm}
\usepackage{bbm}
\usepackage{color}
\usepackage{graphicx}
\usepackage{makecell}
\usepackage{multirow}

\usepackage{hyperref}
\usepackage{dcolumn}
\usepackage{braket}
\usepackage{autobreak}
\usepackage{ulem}
\usepackage{xspace}
\usepackage{tikz}
\usetikzlibrary{quantikz}
\usetikzlibrary{external}
\newcommand{\SUM}{\mathrm{SUM}}
\newcommand{\SWAP}{\mathrm{SWAP}}
\newcommand{\SZ}{\mathit{SZ}}
\newcommand{\SX}{\mathit{SX}}
\newcommand{\ratio}{\xi}

\newcommand{\ii}{\mathrm{i}}
\newcommand{\field}{\mathbb{F}\xspace}
\newcommand{\naturals}{\mathbb{N}\xspace}
\newcommand{\integers}{\mathbb{Z}\xspace}

\newcounter{num}
\newcommand{\Rnum}[1]{\setcounter{num}{1} \Roman{num}}
\newcommand{\rnum}[1]{\setcounter{num}{1} \roman{num}}

\begin{document}

\preprint{}

\title{High-Rate Concatenated Quantum Error-Correcting Codes for Qudits} 

\author{Takanori Nishi}
\email{takanori.nishi@riken.jp}
\affiliation{RIKEN Center for Quantum Computing (RQC), Wako, Saitama 351-0198, Japan.}

\author{Hayato Goto}
\affiliation{RIKEN Center for Quantum Computing (RQC), Wako, Saitama 351-0198, Japan.}
\affiliation{Corporate Research \& Development Center, Toshiba Corporation, 1 Komukai Toshiba-cho, Saiwai-ku, Kawasaki-shi 212-8582, Japan.}
%


\begin{abstract} 
Qudits provide a larger local Hilbert space than qubits and have recently attracted interest as building blocks for efficient fault-tolerant quantum computation (FTQC).
In this work, we extend many-hypercube (MHC) codes, high-rate concatenated quantum codes, from qubits to qudits of prime dimension $q$. 
We construct the level-2 and level-3 MHC codes with parameters $\left[\!\left[6^2,4^2,2^2\right]\!\right]_q$ and $\left[\!\left[6^3,4^3,2^3\right]\!\right]_q$, respectively, and generalize the level-by-level minimum-distance (LLMD) decoder originally proposed for the qubit MHC codes. 
In a qudit $X$-error model, we evaluate the code-capacity performance of the qudit MHC codes for dimensions up to $q=13$.
We find that the error threshold increases monotonically with the local dimension $q$, reaching $10.0\%$ for $q=13$, nearly double the value of $5.1\%$ for qubits.
Furthermore, for sufficiently large local dimensions, we observe a pronounced waterfall regime in which the logical error rate decreases substantially faster than expected from conventional distance-based estimates.
To understand the origin of the observed qudit advantage, we analyze the decoding performance in terms of the entropy of the underlying $q$-ary noise channel. 
This analysis reveals a competition between the increased syndrome information available at larger $q$ and the simultaneous growth of the typical physical error weight at fixed entropy. 
These results demonstrate that higher-dimensional quantum systems can substantially improve the performance of high-rate concatenated quantum codes and motivate further development of qudit-based FTQC architectures. 
\end{abstract}


\maketitle 

\section{INTRODUCTION}\label{Sec:Intro}
Quantum error correction (QEC) \cite{Shor1995ShorCode, Steane1996SteaneCode, CalderbankShor1996CSS, Kitaev2003ToricCode, Nielsen_Chuang_2010} is a fundamental ingredient for realizing scalable fault-tolerant quantum computation (FTQC) \cite{Shor1996FTQC, Gottesman1998FTQC, Nielsen_Chuang_2010}.
By encoding quantum information into a larger physical Hilbert space, QEC protects logical states against decoherence, control imperfections, and other noise sources.
While most theoretical and experimental developments in QEC have focused on qubits \cite{BravyiKitaev1998QuantumCodesLatticeBoundary, Google2023ScalingSurfaceCode, Gottesman2014QLDPC, BreuckmannEberhardt2021QLDPC, Bravyi2024BBcode, PaetznickQuantinuum2024DemonstrationRepeatedQEC, PaetznickQuantinuum2026LogicalGates, Reichardt2025FTQCwithNeutralAtom, BluvsteinLukin2026FTQCwithNeutralAtom, Goto2024Many-hypercube}, there has recently been growing interest in QEC with qudits \cite{BullockBrennen2007QuditSurfaceCodes, Watson2015QuditSurfaceCodeDecoder, Watson2015QuditColorCodes, Keppens2025FiveQuditCode, Spencer2026quditLDPC}, namely quantum systems with more than two levels.
Qudit systems arise naturally in various physical platforms including trapped ions \cite{Ringbauer2022IonQudit1Q2Qgates, Low2025Ba13LevelQudit, Low2026Ba25LevelQudit, Shi2025IonQuditMultitone1QGate}, neutral atoms \cite{Burshtein2026AtomQuditProposal, Flament2025AtomQuditBECmomentum}, superconducting circuits \cite{Bianchetti2010TransmonQtrit, Brock2025GKPQudit, Champion2025TransmonQuditMultitone1QGate}, and photonic systems \cite{YamazakiNTT2025QuditOpticalFusion, Shalaev2025PhotonicNetworkingTrappedIonQudits}.
In addition to reducing the number of physical carriers required to realize a Hilbert space of a given dimension, qudits can provide richer syndrome information \cite{Andrist2015QuditThresholdImprovement}, reduced circuit complexity \cite{Shi2025IonQuditMultitone1QGate}, enhanced magic-state distillation \cite{CAB2012MSDquditPrimeDimQRM,Campbell2014PRLquditMSDimproved, Saha2026SublogQuditMSDpuncturedRM, Prakash2025lowoverheadqutritMSD, Cervia2025quditMSD_AsymptoticallyGoodCodes}, and broader opportunities for fault-tolerant logical-gate constructions \cite{Watson2015QuditColorCodes}.
Motivated by these advantages, various families of qudit error-correcting codes have recently been investigated. Examples include qudit versions of surface codes \cite{BullockBrennen2007QuditSurfaceCodes, Watson2015QuditSurfaceCodeDecoder}, color codes \cite{Watson2015QuditColorCodes}, quantum Reed--Muller codes \cite{CAB2012MSDquditPrimeDimQRM, Campbell2014PRLquditMSDimproved}, and, more recently, quantum low-density parity-check (qLDPC) codes \cite{Spencer2026quditLDPC}.
Recent theoretical advances include symmetry-based code constructions, such as unified SU$(q)$-based frameworks establishing correspondences among permutation-invariant codes, constant-excitation bosonic codes, and codes in monolithic spin manifolds \cite{Aydin2026quditCodesConvexGeometry}, and representation-theoretic approaches that derive qudit codes from irreducible representations of SU$(q)$ \cite{Uy2025QuditCodesFromIrrepSUd}.
These developments demonstrate the broad potential of qudits for QEC.

While many existing quantum code families focus on improving code distances or error thresholds, high-rate quantum codes \cite{Gottesman2014QLDPC, BreuckmannEberhardt2021QLDPC, Bravyi2024BBcode,YamasakiKoashi2024ConcatenatedHammingCode, Goto2024Many-hypercube} offer a complementary approach by increasing the amount of logical information encoded per physical carrier. Among such codes, many-hypercube (MHC) codes introduced in Ref.~\cite{Goto2024Many-hypercube} have a simple code structure based on concatenation of quantum error-detecting codes.
When the $\left[\!\left[6,4,2\right]\!\right]$ code is chosen as a base code, the MHC code at concatenation level $L$ has parameters $\left[\!\left[6^L,4^L,2^L\right]\!\right]$, thereby encoding an exponentially growing logical Hilbert space while maintaining a simple recursive structure. 
The MHC code family supports an efficient recursive decoder and fault-tolerant logical-gate schemes \cite{Goto2024Many-hypercube}.
However, MHC codes have so far been studied only for qubits. 

In this work, we extend the MHC code family from qubits to qudits of prime dimension $q$, where $q$ is referred to as the local dimension.
We note that the present study focuses on prime dimensions up to $q=13$.
Recent experimental progress in trapped-ion platforms has already demonstrated coherent control and single-shot readout of 13-dimensional qudits \cite{Low2025Ba13LevelQudit}.
We consider two decoding strategies. 
The first is a hard-decision (HD) decoder inspired by Knill's concatenated decoding scheme \cite{Knill2005, Goto2024Many-hypercube}.
The second is a level-by-level minimum-distance (LLMD) decoder originally developed for the qubit MHC codes \cite{Goto2024Many-hypercube}. 
We generalize both the decoders from qubits to qudits.
Using large-scale stabilizer simulations in a qudit $X$-error model, we evaluate the code-capacity performance of qudit MHC codes for several local dimensions.
We find that the performance of the LLMD decoder improves with increasing $q$, with the threshold increasing monotonically and nearly doubling from 5.1\% for $q=2$ to 10.0\% for $q=13$.
Previous studies of qudit topological codes have reported that the threshold generally increases with $q$ \cite{Andrist2015QuditThresholdImprovement}. Our results show that a similar enhancement occurs in a high-rate concatenated code family, despite different code structure and decoding strategy.

Recently, a \textit{waterfall} regime of logical-error suppression, where the logical error rate decreases substantially faster than expected from conventional distance-based estimates, has been identified in qubit qLDPC codes using structure-aware neural decoders \cite{Gu2026WaterfallCascadeDecoder}.
For sufficiently large $q$, we observe an analogous waterfall regime in qudit MHC codes.
The appearance of the waterfall regime indicates that the LLMD decoder strongly suppresses classes of high-weight logical failures, leading to an effective scaling exponent larger than that predicted solely from the code distance. 
To understand the origin of these phenomena, we analyze the results in terms of the entropy of the underlying $q$-ary noise channel.
Our analysis reveals a competition between two effects. 
On one hand, increasing $q$ enriches the syndrome alphabet and improves the distinguishability between logical classes. 
On the other hand, at fixed channel entropy, larger values of $q$ correspond to higher physical error rates and hence larger typical error weights. 
We show that this competition naturally explains the different $q$ dependences in the waterfall and low-error regimes, providing a unified interpretation of the threshold enhancement observed at larger local dimensions.

The remainder of this paper is organized as follows.
In Sec. \ref{Sec:Theory}, we introduce qudit MHC codes, logical gate constructions, and decoding methods. 
In Sec. \ref{Sec:Results}, we present numerical simulation results and discuss the code-capacity performance as a function of the local dimension $q$. 
Finally, in Sec. \ref{Sec:Conclusion}, we summarize the results and discuss future directions.

\section{Method}\label{Sec:Theory}
\subsection{Gates and measurements for qudits}
First, we summarize Pauli gates, Clifford gates, and measurements for qudits. 
A qudit is a $q$-dimensional quantum system and can be defined for arbitrary integer dimensions $q\ge 2$. 
We adopt the modular-qudit formalism, in which the Pauli operators are defined as
\begin{align}
    X = \sum_{n=0}^{q-1} \ket{n+1}\bra{n},\, 
    Z = \sum_{n=0}^{q-1} \omega^n \ket{n}\bra{n},\label{eq:XandZgate}
\end{align}
where $\omega=\exp(\ii2\pi/q)$ and their commutation relations are given by
\begin{align}
    (X^a Z^b)(X^c Z^d) = \omega^{bc-ad}(X^c Z^d)(X^a Z^b), \label{eq:XZcommutation}
\end{align}
and the Pauli group is generated as $\mathcal{P}=\{\omega^cX^aZ^b| a,b,c\in \integers_q\}$.
The Pauli group for $n$ qudits, $\mathcal{P}^n$, is the $n$-fold tensor product of $\mathcal{P}$.
This construction is well defined for arbitrary dimensions $q$, however, basic arithmetic operations can be defined
over a finite field $\field_q$ if and only if $q=p^r$ is a power of a prime for a prime number $p$ and a positive integer $r\in\naturals$.
In the present study, we further restrict our attention to the case $r=1$, which simplifies the code construction because $\field_q\cong \integers_q$.
For $r>1$, such an isomorphism no longer holds and a different definition of the Pauli gates called the Galois qudit \cite{Wills2026ReviewGaloisQudits} is more suitable.
Extending the present construction to Galois qudits is left for future work. 

The Clifford group for $n$ qudits is defined as $\mathcal{C}^n=\{U|U\mathcal{P}^nU^{\dagger}=\mathcal{P}^n\}$.
The Clifford group can be generated by the Fourier gate $F$, the phase gate $S$, and the SUM gate, which are qudit generalizations of the Hadamard, phase, and CNOT gates, respectively:
\begin{align}
    F &= \frac{1}{\sqrt{q}} \sum_{n,m=0}^{q-1} \omega^{mn}\ket{m}\bra{n}, \label{eq:Fgate} \\
    S &= 
    \begin{cases}
        \sum_{n=0}^{q-1}\omega^{n^2/2}\ket{n}\bra{n} \, (q\, \mathrm{ is\, even})\\
        \sum_{n=0}^{q-1}\omega^{n(n-1)/2}\ket{n}\bra{n} \, (q\, \mathrm{ is\, odd}) \\
    \end{cases}, \label{eq:Sgate} \\
    \SUM_{c,t} &=\sum_{n,m=0}^{q-1}\ket{n}_c\ket{m+n}_t\bra{n}_c\bra{m}_t \notag\\
    &=\sum_{n=0}^{q-1}\ket{n}_c\bra{n}_c\otimes X_t^n,\label{eq:SUMgate}
\end{align}
where $c$ and $t$ denote the control and target qudits, respectively.
The following relations are used to construct stabilizers, encoding circuits, logical Clifford gates, and the error-correction circuits,
\begin{align}
    &FX^a F^{\dagger} =Z^a,\, FZ^a F^{\dagger} =X^{-a}, \label{eq:FXF_FZF} \\
    &\SUM_{c,t}X_c^a\otimes X_t^b \SUM^{\dagger}_{c,t} =X_c^a\otimes X_t^{a+b}, \\
    &\SUM_{c,t}Z_c^a\otimes Z_t^b\SUM^{\dagger}_{c,t} =Z_c^{a-b}\otimes Z_t^b.
\end{align}

The $n$-th eigenstate of $X$ is given by $\ket{X_n}\equiv F^{\dagger}\ket{n}=\frac{1}{\sqrt{q}}\sum_m \omega^{-mn}\ket{m}$, whose eigenvalue is $\omega^n$ as $X\ket{X_n}=\frac{1}{\sqrt{q}}\sum_m \omega^{-mn}\ket{m+1}=\omega^n\frac{1}{\sqrt{q}}\sum_m \omega^{-(m+1)n}\ket{m+1}=\omega^n\ket{X_n}$.

While a qubit measurement yields a binary outcome in $\integers_2$, a qudit measurement in the $Z$ basis yields an outcome in $\integers_q$.
An $X$-basis measurement can be implemented by applying the $F$ gate followed by a $Z$-basis measurement.

\subsection{MHC codes for qudits}
MHC codes are obtained by concatenating high-rate error-detecting codes.
The encoding rate is defined as $R=k/n$ for a $\left[\!\left[n,k,d\right]\!\right]_q$ code, where $n$ is the number of physical qudits, $k$ is the number of logical qudits, and $d$ is the code distance.
Such a code encodes a logical Hilbert space of dimension $q^k$ into $n$ physical qudits.
Here, we concatenate the $\left[\!\left[6,4,2\right]\!\right]_q$ code to obtain the level-$L$ MHC code $\left[\!\left[n^{(L)},k^{(L)},d^{(L)}\right]\!\right]_q=\left[\!\left[6^L,4^L,2^L\right]\!\right]_q$.
At level 1, the stabilizer generators are chosen as
\begin{align}
    \SZ^{(1)}=Z_1 Z_2 Z_3 Z_4 Z_5 Z_6,\,
    \SX^{(1)}=X_1 X_2^{\dagger} X_3 X_4^{\dagger} X_5 X_6^{\dagger},
\end{align}
and the logical Pauli gates are defined as 
\begin{align}
    Z^{(1)}_1&=Z_1 Z_2,\, X^{(1)}_1=X_2 X^{\dagger}_3, \notag \\
    Z^{(1)}_2&=Z_2 Z_3,\, X^{(1)}_2=X^{\dagger}_1 X_2, \notag \\
    Z^{(1)}_3&=Z_4 Z_5,\, X^{(1)}_3=X_5 X^{\dagger}_6, \notag \\
    Z^{(1)}_4&=Z_5 Z_6,\, X^{(1)}_4=X^{\dagger}_4 X_5, \notag
\end{align}
where the superscript represents the concatenation level.

\begin{figure}[tb]
\centering
\includegraphics[width=\columnwidth]{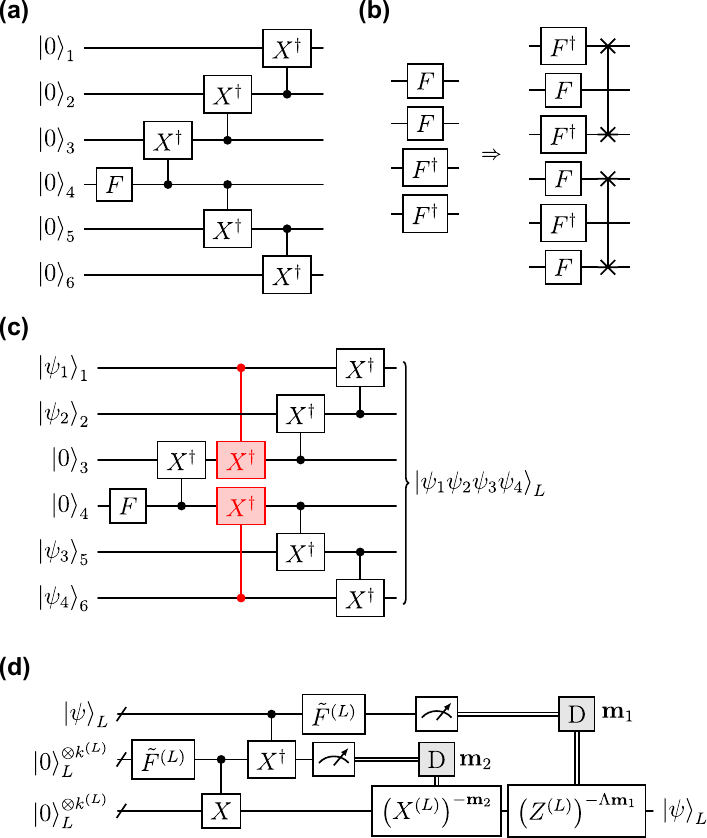}
\caption{
(a) Zero-state encoder of the $\left[\!\left[6,4,2\right]\!\right]_q$ code. (b) Level-1 logical Fourier gate. (c) Arbitrary-state encoder, where additional $\SUM^{\dagger}$ gates are highlighted in red. (d) Error-correcting teleportation for qudit MHC codes.
}
\label{fig:circuits}
\end{figure}

The non-fault-tolerant encoder shown in Fig.~\ref{fig:circuits} (a), consisting of a Fourier gate and five $\SUM^{\dagger}$ gates, prepares the logical zero state
\begin{align}
    \ket{0,0,0,0}_{L1}=\frac{1}{\sqrt{q}}\sum_{n=0}^{q-1} \ket{-n, n, -n, n, -n, n}. \label{eq:logical0state}
\end{align}
A level-1 computational basis state $\ket{a_1,a_2,a_3,a_4}_{L1}$ is stabilized by $\braket{\SZ,\SX,\omega^{-a_1}Z^{(1)}_1, \omega^{-a_2}Z^{(1)}_2, \omega^{-a_3}Z^{(1)}_3, \omega^{-a_4}Z^{(1)}_4}$ for $a_j\in \integers_q$.

The fully transversal logical SUM gate is obtained by the transversal application of physical SUM gates as
\begin{align}
    \bigotimes_{j'=1}^{4}\SUM^{(1)}_{c_{j'},t_{j'}}=\bigotimes_{j=1}^{6}\SUM_{c_j,t_j},
\end{align}
where $c_j$ and $t_j$ denote the $j$-th qudit in the control and target block, respectively.

The logical Fourier gate, however, cannot be fully transversal.
Instead, by requiring Eq. \eqref{eq:FXF_FZF} to hold at the logical level for all logical Pauli gates, we can obtain the following logical gate,
\begin{align}
\label{eq:tildeF(1)}
    \tilde{F}^{(1)}=F^{(1)}_1 F^{(1)}_2 F^{(1)\dagger}_3 F^{(1)\dagger}_4,
\end{align}
which can be implemented using $F$, $F^{\dagger}$, and SWAP gates as shown in Fig. \ref{fig:circuits} (b). 

To illustrate higher-level cases, we consider $L=3$.
Similarly to the qubit MHC codes \cite{Goto2024Many-hypercube}, we label the physical qudits using three integers as $\{Q_{i,j,k} | i,j,k=1,\dots,6\}$ and encode the four level-1 logical qudits into the six physical qudits for each block labeled by $(j,k)$.
The obtained blocks of the level-1 qudits are labeled as $\{Q^{(1)}_{i',j,k} | i'=1,\dots,4\,,j,k=1,\dots,6\}$.
The stabilizer generators and the Pauli operators can be labeled as
\begin{align}
    \SZ^{(1)}_{j,k}&=\bigotimes_{i=1}^6 Z_{i,j,k}, \,
    \SX^{(1)}_{j,k}=\bigotimes_{i=1}^6 X_{i,j,k}^{\sigma_i}, \\
    
    Z^{(1)}_{1,j,k}&=Z_{1,j,k}Z_{2,j,k}, \, X^{(1)}_{1,j,k}=X_{2,j,k}X^{\dagger}_{3,j,k}, \\
    
    Z^{(1)}_{2,j,k}&=Z_{2,j,k}Z_{3,j,k}, \, X^{(1)}_{2,j,k}=X^{\dagger}_{1,j,k}X_{2,j,k}, \\
    
    Z^{(1)}_{3,j,k}&=Z_{4,j,k}Z_{5,j,k}, \, X^{(1)}_{3,j,k}=X_{5,j,k}X^{\dagger}_{6,j,k}, \\
    
    Z^{(1)}_{4,j,k}&=Z_{5,j,k}Z_{6,j,k}, \, X^{(1)}_{4,j,k}=X^{\dagger}_{4,j,k}X_{5,j,k},
\end{align}
where $\sigma_i=(-1)^{i+1}$.
Then, the four level-2 qudits are encoded into the six level-1 qudits for each block labeled by $(i',k)$ and the obtained level-2 blocks are labeled as $\{Q^{(2)}_{i',j',k} | i',j'=1,\dots,4\,,k=1,\dots,6\}$.
Similarly, the four level-3 qudits are encoded into the six level-2 qudits for each block labeled by $(i',j')$ and the obtained level-3 block is labeled as $\{Q^{(3)}_{i',j',k'} | i',j',k'=1,\dots,4\}$.
The stabilizers $\{\SZ_{i',j'}^{(3)},\,\SX_{i',j'}^{(3)}|i',j'=1,\dots,4\}$ and the Pauli operators $\{Z^{(3)}_{i',j',k'},\,X^{(3)}_{i',j',k'}|i',j',k'=1,\dots,4\}$ can be derived by decomposing Pauli operators recursively. For example,
\begin{align}
    Z^{(3)}_{1,1,1}&= Z^{(2)}_{1,1,1}Z^{(2)}_{1,1,2}=Z^{(1)}_{1,1,1}Z^{(1)}_{1,2,1}Z^{(1)}_{1,1,2}Z^{(1)}_{1,2,2} \notag \\
                   &=Z_{1,1,1}Z_{2,1,1}Z_{1,2,1}Z_{2,2,1}Z_{1,1,2}Z_{2,1,2}Z_{1,2,2}Z_{2,2,2},\\
    X^{(3)}_{1,1,1}&= X^{(2)}_{1,1,2}X^{(2)\dagger}_{1,1,3}=X^{(1)}_{1,2,2}X^{(1)\dagger}_{1,3,2}X^{(1)\dagger}_{1,2,3}X^{(1)}_{1,3,3} \notag \\
                   &=X_{2,2,2}X^{\dagger}_{3,2,2}X^{\dagger}_{2,3,2}X_{3,3,2}X^{\dagger}_{2,2,3}X_{3,2,3}X_{2,3,3}X^{\dagger}_{3,3,3},\\
    \SZ^{(3)}_{3,2}&=\bigotimes_{k=1}^6 Z^{(2)}_{3,2,k}=\bigotimes_{k=1}^6 Z^{(1)}_{3,2,k}Z^{(1)}_{3,3,k} \notag\\
                   &=\bigotimes_{k=1}^6 Z_{4,2,k}Z_{5,2,k}Z_{4,3,k}Z_{5,3,k}, \\
    \SX^{(3)}_{3,1}&=\bigotimes_{k=1}^6 \left(X^{(2)}_{3,1,k}\right)^{\sigma_k}=\bigotimes_{k=1}^6 \left(X^{(1)}_{3,2,k}\right)^{\sigma_k} \left(X^{(1)}_{3,3,k}\right)^{-\sigma_k} \notag\\
                   &=\bigotimes_{k=1}^6 X^{\sigma_k}_{5,2,k}X^{-\sigma_k}_{6,2,k}X^{-\sigma_k}_{5,3,k}X^{\sigma_k}_{6,3,k}.             
\end{align}

The fully transversal logical SUM gate is obtained by applying the physical SUM gates between physical qudits as
\begin{align}
    \bigotimes_{i',j',k'=1}^{4}\SUM^{(3)}_{c_{i',j',k'},t_{i',j',k'}}=\bigotimes_{i,j,k=1}^{6}\SUM_{c_{i,j,k},t_{i,j,k}}.
\end{align}
The logical $F$ gate can be constructed by using Eq. \ref{eq:tildeF(1)} recursively as
\begin{align}
    \tilde{F}^{(3)}=\bigotimes_{i',j',k'=1}^4 \left(F^{(3)}_{i',j',k'}\right)^{\lambda_{i'}\lambda_{j'}\lambda_{k'}},
\end{align}
where 
\begin{align}
\label{eq:lambda_for_Fourier}
\lambda_{a'}=
\begin{cases}
1  & a'\in\{1,2\}\\
-1 & a'\in\{3,4\}
\end{cases}.
\end{align}
By applying the map shown in Fig. \ref{fig:circuits} (b) recursively, $\tilde{F}^{(3)}$ can be implemented as
\begin{align}
    &\bigotimes_{i,j=1}^6\left[ \SWAP_{(i,j,1)\longleftrightarrow(i,j,3)}\SWAP_{(i,j,4)\longleftrightarrow(i,j,6)} \right] \notag\\
    &\bigotimes_{i,k=1}^6\left[ \SWAP_{(i,1,k)\longleftrightarrow(i,3,k)}\SWAP_{(i,4,k)\longleftrightarrow(i,6,k)} \right] \notag\\
    &\bigotimes_{j,k=1}^6\left[\SWAP_{(1,j,k)\longleftrightarrow(3,j,k)}\SWAP_{(4,j,k)\longleftrightarrow(6,j,k)}\right] \notag\\
    &\bigotimes_{i,j,k=1}^6 \left(F_{i,j,k}\right)^{\eta_i\eta_j\eta_k} 
    ,
\end{align}
where $\eta_a=(-1)^a$.
The logical Fourier gates at higher levels can be defined in a similar way.

Non-Clifford gates can be implemented by gate teleportation using distilled magic states. 
Noisy encoded magic states can be prepared using an arbitrary-state encoder shown in Fig.~\ref{fig:circuits} (c) and subsequently distilled via qudit Reed--Muller codes \cite{CAB2012MSDquditPrimeDimQRM, Campbell2014PRLquditMSDimproved, Saha2026SublogQuditMSDpuncturedRM} or other qudit triorthogonal codes \cite{Prakash2025lowoverheadqutritMSD, Cervia2025quditMSD_AsymptoticallyGoodCodes}.

\subsection{Decoders} \label{subsec:decoders}

To perform error correction, we adopt error-correcting teleportation (ECT) \cite{Knill2005, Goto2024Many-hypercube}, shown in Fig.~\ref{fig:circuits} (d).
The decoder receives the measurement outcomes of all physical qudits in the computational basis and estimates the logical Bell-measurement results.
We denote the outcomes of the decoders by $\mathbf{m}_1\in \integers_q^{k^{(L)}}$ and $\mathbf{m}_2 \in \integers_q^{k^{(L)}}$, respectively, whose elements are labeled by $L$ indices, e.g., $(\mathbf{m}_1)_{i',j',k'}\, (i',j',k'=1,\dots,4)$ at level 3.
The correction operator $X^{(L)}$ must be raised to the power of $-\mathbf{m}_2$, which is a tensor product of the logical operators, e.g., $\left(X^{(3)}\right)^{-\mathbf{m}_2}=\bigotimes_{i',j',k'=1}^4\left(X^{(3)}_{i',j',k'}\right)^{-(\mathbf{m}_{2})_{i',j',k'}}$ for $L=3$.
Similarly, the correction operator $Z^{(L)}$ must be raised to the power of $-\Lambda\mathbf{m}_1$, where $\Lambda$ is the diagonal matrix whose diagonal elements are labeled by $L$ indices, e.g., $\Lambda_{i',j',k'}=\lambda_{i'}\lambda_{j'}\lambda_{k'} \,(i',j',k'=1,\dots,4)$  for $L=3$, where $\lambda_{a'}$ is defined by Eq. \ref{eq:lambda_for_Fourier}.

As introduced in Sec.~\ref{Sec:Intro}, we consider the HD and LLMD decoders, both of which are qudit extensions of the qubit decoders introduced in Ref.~\cite{Goto2024Many-hypercube}.

The HD decoder recursively checks stabilizer consistency at each concatenation level. 
A block is accepted if its syndrome is consistent with the code constraints and is otherwise flagged as a failure.
Higher-level blocks are then decoded using the available lower-level information and discarded whenever multiple incompatible failures occur.
The HD decoder has a low computational cost but does not explicitly search for the most likely logical state. 
The details are provided in Appendix~\ref{App:HDdecoder}.

In contrast, the LLMD decoder retains all encoded strings that achieve the minimum Hamming distance from the observed measurement outcomes. These candidates are propagated recursively through the concatenation hierarchy while enforcing the stabilizer constraints between neighboring blocks. The final logical state is chosen from the minimum-distance candidates at the highest level. 
This procedure approximates global minimum-distance decoding without an exponential search over the entire code space.
Detailed decoding procedures, candidate-generation rules, distance calculations, and complexity-reduction techniques are provided in Appendix~\ref{App:LLMDdecoder}.

\section{Results} \label{Sec:Results}
\subsection{Code-capacity simulations} \label{subsec:code_capacity}

\begin{figure*}
\centering
\includegraphics[width=\textwidth]{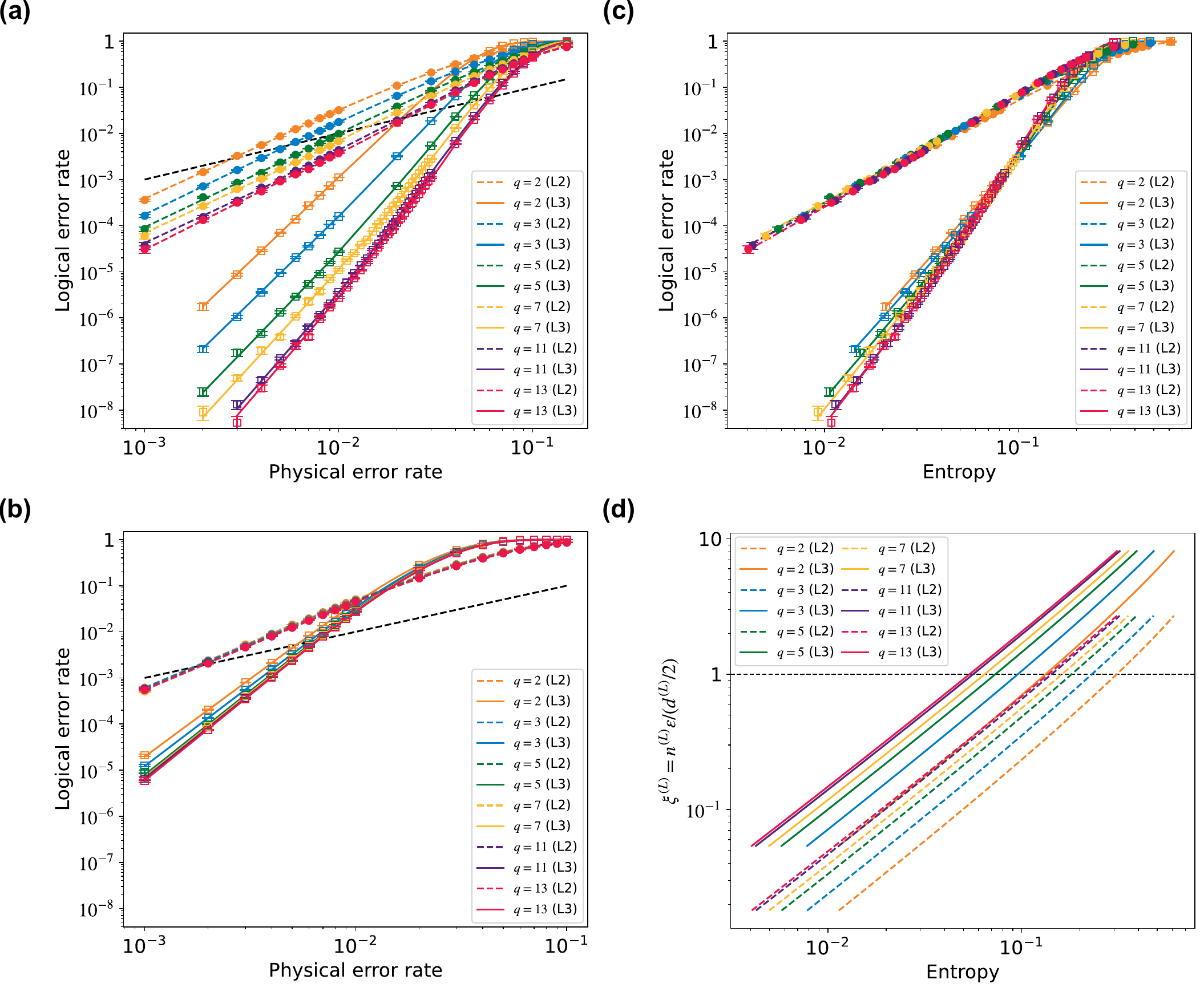}
\caption{
Code-capacity performance, entropy scaling, and typical-error-weight analysis of qudit MHC codes under the $X$-error model.
(a) Code-capacity performance obtained with the LLMD decoder. (b) Code-capacity performance obtained with the HD decoder. (c) Logical error rate for the LLMD decoder replotted as a function of the $q$-ary entropy $H_q$ using the same data as in (a). (d) Ratio of the typical error weight to one-half of the code distance as a function of $H_q$. Colors indicate the local dimension $q$. In (a)--(d), dashed and solid curves correspond to the level-2 and level-3 concatenation, respectively. In (a) and (b), thresholds are defined by the crossing points of the level-2 and level-3 curves, and the black dashed line indicates the break-even condition, $\varepsilon_L=\varepsilon$. The horizontal dashed line in (d) marks the point at which the typical error weight equals one-half of the code distance. 
Error bars in (a)--(c) indicate one standard deviation estimated from Monte Carlo sampling.
At level 2, $10^4$ and $10^4$--$10^6$ shots were used for the HD and LLMD decoders, respectively. 
At level 3, $10^6$--$2\times10^7$ and $10^4$--$1.5\times10^9$ shots were used for the HD and LLMD decoders, respectively, with larger sample sizes used at larger $q$ and lower physical error rates.}
\label{fig:capacity_entropy}
\end{figure*}
We numerically evaluated the code capacity of qudit MHC codes under a qudit $X$-error model.
As in the qubit case, Clifford circuits on qudits can be efficiently simulated classically within the stabilizer formalism.
In particular, we adopt the stabilizer tableau simulator for qudits called \texttt{Sdim} \cite{kabir2026Sdim}.
First, we prepare an error-free logical zero state and then each physical qudit is independently subject to $X$-type errors with probability $\varepsilon$.
More precisely, we assume that the probability of an $X^a$ error is independent of $a$ and the noise channel assumed here is defined as $\mathcal{E}(\rho)=(1-\varepsilon)\rho + \frac{\varepsilon}{q-1}\sum_{a=1}^{q-1} X^a \rho X^{-a}$.
Then, all the physical qudits are measured without errors and the measurement outcomes are decoded by the HD or LLMD decoder.
We define a decoding trial to be successful only when all the logical qudits are decoded correctly, that is, all the logical-qudit values estimated by the decoding are zero.
This definition corresponds to a block error rate rather than a per-logical-qudit error rate.
We repeat the above procedure sufficiently many times to reduce the statistical error.
The results obtained with the LLMD and HD decoders are shown in Fig.~\ref{fig:capacity_entropy} (a) and (b), respectively.

\begin{table*}[t]
\caption{Fitted exponents and estimated thresholds from the data in Figs. \ref{fig:capacity_entropy} (a) and \ref{fig:capacity_entropy} (b).
The range of $\varepsilon$ used for the fit is shown in the parentheses in the second column.
}
\label{tab:threshold_exponent}
\centering
\setlength{\tabcolsep}{10pt}
\begin{tabular}{clcccccc}
\hline\hline

\multirow{2}{*}{Decoder}
&
\multirow{2}{*}{}
&
\multicolumn{6}{c}{$q$}
\\

&
&
2 & 3 & 5 & 7 & 11 & 13
\\

\hline

\multirow[t]{4}{*}{LLMD}
& Level 2 (0.001--0.005)
& 1.94(2) & 2.03(2) & 2.00(10) & 1.97(6) & 1.97(10) & 2.12(4)
\\

& Level 3 (0.003--0.01)
& 4.01(3) & 4.09(3) & 4.35(5) & 4.58(7) & 4.80(6) & 4.90(8)
\\

& Level 3 (0.02--0.04)
& 3.48(9) & 4.30(1) & 5.07(3) & 5.35(2) & 5.59(3) & 5.73(2)
\\

& Threshold (\%)
& 5.1 & 7.3 & 8.5 & 9.1 & 9.8 & 10.0
\\

\hline

\multirow[t]{4}{*}{HD}
& Level 2 (0.001--0.005)
& 1.95(2) & 1.90(3) & 1.96(2) & 1.94(1) & 1.92(2) & 1.92(2)
\\

& Level 3 (0.001--0.005)
& 3.38(0) & 3.48(1) & 3.63(1) & 3.65(1) & 3.65(2) & 3.72(1)
\\

& Threshold (\%)
& 1.1 & 1.3 & 1.4 & 1.4 & 1.4 & 1.4
\\

\hline\hline

\end{tabular}
\end{table*}

For both decoders, the thresholds summarized in Table \ref{tab:threshold_exponent} are defined as the crossing points of the level-2 and level-3 curves.
As shown in Table \ref{tab:threshold_exponent}, the threshold monotonically increases with $q$ for the LLMD decoder, reaching 10.0\% for $q=13$, nearly double the qubit threshold of 5.1\%.
In contrast, the $q$-dependence is not clear for the HD decoder.

To characterize the scaling behavior of the logical error rate, we fit the data to the power law $\varepsilon_L \propto \varepsilon^\alpha$. 
The fitted exponents are summarized in Table~\ref{tab:threshold_exponent}. 
For the HD decoder, the exponent exhibits only a weak dependence on $q$ for both concatenation levels. 
In contrast, the LLMD decoder shows a clear improvement with increasing $q$.
While the fitted exponents for the level-2 code remain nearly independent of $q$, those for the level-3 code increase systematically as $q$ increases.
Moreover, the level-3 code exhibits distinct scaling behaviors in the low- and high-error regions.
In the high-error region ($\varepsilon$=0.02--0.04), the fitted exponents exceed the value expected from the code distance, $\lfloor (d^{(3)}+1)/2 \rfloor = 4$, reaching values greater than 5 for $q\geq5$. 
By contrast, the fitted exponents in the low-error region ($\varepsilon$=0.003--0.01) remain closer to the distance-based expectation, although they also increase with $q$.
To understand this behavior, we write the logical error rate schematically as $\varepsilon_L=\sum_w p_wN_w \varepsilon^w(1-\varepsilon)^{n^{(L)}-w}$,
where $N_w$ denotes the number of weight-$w$ error configurations and $p_w$ is the probability that the decoder fails on such configurations. 
In conventional distance-based estimates, logical failures are assumed to be dominated by the smallest uncorrectable errors beyond the decoding radius $\lfloor (d^{(L)}-1)/2\rfloor$. 
However, if the decoder significantly reduces $p_w$ for $w$ beyond the decoding radius, the dominant contribution shifts to higher-weight error configurations, leading to a steeper effective scaling of $\varepsilon_L$.
The high-error region exhibiting such a steep drop of the logical error rate is called the waterfall regime \cite{Gu2026WaterfallCascadeDecoder} while the low-error region where the exponent converges to $\lfloor (d^{(L)}+1)/2 \rfloor$ is called the error-floor regime.
The exponents in Table \ref{tab:threshold_exponent} suggest that the LLMD decoder significantly reduces $p_4$ for $q\geq 5$ and $p_5$ as well for $q\geq 7$ and, therefore, the waterfall regime appears for $q\geq 5$.
We will refer to the low-error region as the error-floor regime for convenience although the exponents for $q\geq5$ have not converged.

\subsection{Entropy-based interpretation} \label{subsec:Entropy}

We investigate the mechanism of the $q$-dependent behaviors in terms of entropy of the noise channel.
The $q$-ary entropy of the noise channel used here is given by
\begin{align}
    H_q(\varepsilon)=-(1-\varepsilon)\log_q(1-\varepsilon) - \varepsilon \log_q\left(\frac{\varepsilon}{q-1}\right).
\end{align}
By construction, $H_q$ is normalized so that its maximum value is $H_q=1$, which occurs at $\varepsilon=(q-1)/q$.
For the range of $\varepsilon$ considered here, increasing $q$ leads to a decrease in entropy $H_q$, for which less information is lost and decoding is generally expected to be easier.
In Fig. \ref{fig:capacity_entropy} (c), we replot the logical error rate for the LLMD decoder in Fig. \ref{fig:capacity_entropy} (a) as a function of $H_q$.
At level 2, the logical error rates collapse onto a single curve.
This indicates that the $q$-dependence observed at fixed physical error rate in Fig.~\ref{fig:capacity_entropy} (a) originates from the corresponding change in the entropy of the $q$-ary noise channel.

A similar trend is observed at level 3, where the curves also move closer together when plotted as a function of $H_q$. 
This suggests that the entropy of the underlying channel remains the dominant factor governing the decoding performance. 
However, a dependence on $q$ still remains, indicating that additional effects beyond the channel entropy must be taken into account.
This $q$-dependence can be explained by two competing effects.
First, a larger value of $q$ provides a richer syndrome alphabet, which can improve the distinguishability between competing logical classes. 
Second, at fixed $H_q$, a larger value of $q$ results in a higher physical error rate $\varepsilon$ and, therefore, a larger typical error weight $w^{(L)}\approx n^{(L)}\varepsilon$, which can make decoding more difficult.
In the waterfall regime, the degradation associated with higher-weight error configurations appears to outweigh the benefit of the increased syndrome information available at larger $q$, resulting in a higher logical error rate for larger $q$.
In contrast, the error-floor regime is governed by low-weight error configurations, for which the enhanced syndrome information available at larger $q$ improves the distinguishability of logical classes and suppresses logical errors.

To clarify the relation between the growth of the typical error weight and the code distance, we plot the ratio $\ratio^{(L)}\equiv n^{(L)}\varepsilon/(d^{(L)}/2)$ as a function of $H_q$ since  $\ratio^{(L)}$ quantifies the proximity of the typical error weight to the weight of the lowest-weight uncorrectable error.
As shown in Fig.~\ref{fig:capacity_entropy} (d), $\ratio^{(3)}$ becomes significantly larger than unity in the large-entropy region, whereas $\ratio^{(2)}$ remains closer to one. Moreover, at fixed $H_q$, the ratio $\ratio^{(L)}$ increases with $q$. These observations indicate that, although the entropy of the underlying channel provides the dominant effect for the decoding performance, increasing $q$ also increases the typical error weight relative to the decoding radius. 
This effect is pronounced at level 3 and provides a natural explanation for the residual $q$-dependence observed in Fig.~\ref{fig:capacity_entropy} (c), particularly in the waterfall regime.

\section{Conclusion} \label{Sec:Conclusion}
We have introduced a qudit extension of MHC codes and investigated their code-capacity in a qudit $X$-error model. 
We constructed the level-2 and level-3 qudit MHC codes, $\left[\!\left[6^2,4^2,2^2\right]\!\right]_q$ and $\left[\!\left[6^3,4^3,2^3\right]\!\right]_q$, respectively, together with their stabilizers, logical operators, encoding circuits, logical Clifford gates, error-correcting teleportation protocol, and the corresponding decoding schemes. 
The resulting codes preserve the high encoding rates of the original qubit construction while naturally supporting the logical qudits of local dimension $q$. 

Numerical simulations revealed that the LLMD decoder substantially outperforms the HD decoder and that its threshold increases monotonically with the local dimension, reaching $10.0\%$ for $q=13$, nearly double the qubit threshold of 5.1\%.
These results demonstrate that higher-dimensional quantum systems can significantly improve the performance of high-rate concatenated quantum codes when the available syndrome information is utilized effectively.
More broadly, our results suggest that combining high-rate concatenated code architectures with higher-dimensional quantum systems and advanced decoding techniques may offer a promising route toward low-overhead FTQC.

Another notable observation is the emergence of a waterfall regime for sufficiently large values of $q$.
In this regime, the logical error rate decreases significantly faster than expected from a simple distance-based estimate, indicating that the LLMD decoder suppresses a large fraction of high-weight logical failures. 
The recent observation of a waterfall regime in qubit qLDPC codes using advanced neural decoders has attracted considerable interest \cite{Gu2026WaterfallCascadeDecoder}.
Our results show that a similar phenomenon can also arise in a fundamentally different code family, namely high-rate concatenated qudit codes. 
To understand the origin of the observed qudit advantage, we analyzed the decoding performance as a function of the $q$-ary entropy of the noise channel. 
This analysis reveals a competition between two effects.
Increasing the local dimension enriches the syndrome alphabet and improves the distinguishability between logical classes, whereas at fixed entropy it also increases the typical physical error weight. 
We find that the relative importance of these effects depends on the operating regime, providing a natural explanation for the distinct behavior observed in the waterfall and error-floor regimes.
Importantly, our results show that the richer syndrome information available at larger local dimensions can itself contribute to logical-error suppression, beyond what can be explained solely by entropy-based considerations.

Since the benefits observed in the present study originate in part from the increased information available in higher-dimensional syndrome alphabets, more powerful decoders may exploit this information more effectively and yield even greater improvements as the local dimension increases.
Recently, decoding strategies for concatenated quantum codes have been further developed \cite{Zhang2026BidirectionalDecoding, Wu2026HierarchicalMessagePassing} and the decoding performance of qubit MHC codes has been improved under a circuit-level noise model \cite{Wu2026HierarchicalMessagePassing}.
Developing analogous decoding algorithms for qudit MHC codes is therefore a promising direction for future work.

Our analysis of the waterfall regime relies on the Hamming distance used in the LLMD decoder and on the corresponding notion of error weight. 
The Hamming distance is appropriate for the present noise model because the probabilities of all $X^a$ errors with $a\neq0$ are equal. 
For more structured noise models, however, other distance measures may be more suitable.
For example, in the case of amplitude damping, the distance should take into account the relaxation pathways and transition probabilities restricted by selection rules, which depend on physical systems such as superconducting circuits, trapped ions, neutral atoms, and optical photons.
Noise-adapted qudit codes \cite{Grassl2018CodeForQuditAmplitudeDamping} and recovery schemes  \cite{Dutta2025QuditCodesForAmpliduteDamping} have been proposed for amplitude-damping channels, where the dominant error processes correspond to asymmetric transitions between energy levels rather than symmetric Pauli errors.
We leave the extension of the LLMD decoder to such noise models for future work.

Our choice of $q\le 13$ is not intended to represent a fundamental experimental limit. 
Rather, it reflects a compromise between computational tractability and experimental relevance. The dimensions considered here are sufficient to reveal clear trends in the threshold, waterfall behavior, and entropy scaling, while remaining directly connected to currently accessible qudit hardware \cite{Low2025Ba13LevelQudit}. 
Exploring still larger dimensions, including experimentally demonstrated $q=25$ systems \cite{Low2026Ba25LevelQudit}, is an interesting direction for future work.
Additional future directions include extending the construction to Galois qudits with dimensions $q=p^r$, investigating circuit-level noise, and developing noise-adapted decoding strategies for physically relevant non-Pauli noise channels such as amplitude damping.

\begin{acknowledgments}
This work was supported by JST-CREST Grant No. JPMJCR23I7 and JST Moonshot R\&D Grant
Number JPMJMS256E.
Parts of
numerical simulations were done on the HOKUSAI supercomputer at RIKEN (project ID RB230022).
\end{acknowledgments}

\appendix 

\section{Decoder details} \label{App:Decoders}
Here, we describe the decoding procedures for the level-3 MHC code. The procedures for the level-2 MHC code are analogous. 
We refer to the output of the decoder at level~$l$ as \textit{a level-$l$ encoded string} $\mathbf{x}^{(l)}\in\integers_q^{4^l}$. 
We also refer to a set of six level-$(l-1)$ encoded strings, or six physical strings for $l=1$, that constitute the level-$l$ encoded string as \textit{a level-$l$ block}. 
For level-3 decoding, we denote the physical-qudit measurement outcomes by $\mathbf{x}=\{x_{i,j,k}\in \integers_q\mid i,j,k=1,\dots,6\}$, the level-1 encoded strings by $\mathbf{x}^{(1)}_{j,k}=\left(x^{(1)}_{1,j,k},\dots,x^{(1)}_{4,j,k}\right)$, and the level-2 encoded strings by $\mathbf{x}^{(2)}_{k}=\left(x^{(2)}_{1,1,k},\dots,x^{(2)}_{4,1,k}, x^{(2)}_{1,2,k},\dots,x^{(2)}_{4,4,k}\right)$.

\subsection{HD decoder}\label{App:HDdecoder}
HD decoding was originally proposed by Knill for the $C_4/C_6$ scheme \cite{Knill2005} and extended to the qubit MHC code by Goto \cite{Goto2024Many-hypercube}. 
For each level-1 block labeled by $(j,k)$, we calculate the syndrome $s^{(1)}_{j,k}=\sum_{i=1}^6 x_{i,j,k}$. 
If $s^{(1)}_{j,k}\neq0$, the level-1 encoded string $\mathbf{x}^{(1)}_{j,k}$ is set to $F$, where $F$ is a flag indicating a level-1 decoding failure. 
Otherwise, the level-1 encoded string is obtained from the definitions of $Z^{(1)}_{i',j,k}$ as  $x^{(1)}_{1,j,k}=x_{1,j,k}+x_{2,j,k}$, $x^{(1)}_{2,j,k}=x_{2,j,k}+x_{3,j,k}$,  $x^{(1)}_{3,j,k}=x_{4,j,k}+x_{5,j,k}$, and $x^{(1)}_{4,j,k}=x_{5,j,k}+x_{6,j,k}.$
Next, for each level-2 block labeled by $k$, the level-2 encoded string $\mathbf{x}^{(2)}_k$ is obtained from the six level-1 encoded strings $\{\mathbf{x}^{(1)}_{j,k}\mid j=1,\dots,6\}$ as follows. 
If exactly one of the six strings, e.g., $\mathbf{x}^{(1)}_{1,k}$, is $F$, we recover it by imposing the stabilizer condition $\mathbf{x}^{(1)}_{1,k}=-\sum_{j=2}^6\mathbf{x}^{(1)}_{j,k}.$ 
The level-2 encoded string $\mathbf{x}^{(2)}_k$ is then obtained from the definitions of $Z^{(2)}_{i',j',k}$. 
If none of the six level-1 strings is $F$, the syndrome vector is calculated as $\mathbf{s}^{(2)}_k=\sum_{j=1}^6\mathbf{x}^{(1)}_{j,k}$. 
If $s^{(2)}_{i',k}=0$ for all $i'$, the level-2 encoded string $\mathbf{x}^{(2)}_k$ is obtained from the definitions of $Z^{(2)}_{i',j',k}$. 
Otherwise, it is set to $F$. 
If two or more level-1 strings are $F$, $\mathbf{x}^{(2)}_k$ is also set to $F$. The level-3 encoded string $\mathbf{x}^{(3)}$ is obtained similarly.
If level-3 decoding fails, each component of the decoded string is assigned a uniformly random value from $\integers_q$.

\subsection{LLMD decoder} \label{App:LLMDdecoder}
Minimum-distance decoding searches for a codeword that minimizes the Hamming distance from the measurement outcomes. 
Although it can provide high decoding performance, its computational cost is high because the number of codewords scales exponentially as $O(q^k)$ for a $\left[\!\left[n,k,d\right]\!\right]_q$ code. 
To reduce the number of codewords that must be examined, the LLMD decoder for qubits proposed in Ref.~\cite{Goto2024Many-hypercube} retains only the minimum-distance codewords at each concatenation level. 
For qudit MHC codes, we also use the Hamming distance $d_H$. 
That is, we count only the number of positions at which the two strings differ, without considering the magnitudes of their differences. 
For example,  $d_H(x_0,x_1)=d_H(x_0,x_2)=2$ for $x_0=(0,0,0,0,0,0)$, $x_1=(1,0,1,0,0,0)$, and $x_2=(2,0,0,0,3,0)$.

For the level-3 MHC codes, we first calculate the syndrome $s^{(1)}_{j,k}$ for each level-1 block labeled by $(j,k)$. 
We denote the measurement outcomes in this block by $\mathbf{x}_{j,k}=(x_{1,j,k},\dots,x_{6,j,k})$.
If $s^{(1)}_{j,k}=0$, $\mathbf{x}_{j,k}$ is a minimum-distance codeword with $d_{j,k} = d_H(\mathbf{x}_{j,k}, \mathbf{x}_{j,k})=0$, and the corresponding level-1 encoded string $\mathbf{x}^{(1)}_{j,k}$ is determined from the definitions of $Z^{(1)}_{i',j,k}$. 
If $s^{(1)}_{j,k}\neq0$, there are six minimum-distance codewords, each at Hamming distance one from $\mathbf{x}_{j,k}$. 
They are given by $\{\mathbf{x}_{j,k}-s^{(1)}_{j,k}\mathbf{e}_i\mid i=1,\dots,6\}$, where $\mathbf{e}_i\in\integers_2^6$ and $(\mathbf{e}_i)_r=\delta_{i,r}$.
For each minimum-distance codeword, a level-1 encoded string $\mathbf{x}^{(1)}_{j,k}$ is obtained from the definitions of $Z^{(1)}_{i',j,k}$ and retained as a minimum-distance candidate.

Next, for each level-2 block labeled by $k$, we choose candidates for five level-1 encoded strings, e.g., $\{\mathbf{x}^{(1)}_{j,k}\mid j=1,\dots,5\}$, and determine the remaining encoded string by imposing the stabilizer condition $\mathbf{x}^{(1)}_{6,k} = -\sum_{j=1}^5\mathbf{x}^{(1)}_{j,k}$.
Because the resulting $\mathbf{x}^{(1)}_{6,k}$ may not be included in the minimum-distance candidates obtained in the previous step, we calculate  its distance from the corresponding measurement outcomes as 
\begin{align} d' = \min_{\mathbf{y}\in\widetilde{\mathcal{X}}} d_H\left(\mathbf{x}_{6,k} ,\mathbf{y} \right), \notag
\end{align}
where $\widetilde{\mathcal{X}}$ is the set of the $q$ length-6 computational-basis strings corresponding to $\mathbf{x}^{(1)}_{6,k}$, as illustrated for the level-1 logical zero state in Eq.~\eqref{eq:logical0state}.
Adding the distances associated with the other five level-1 encoded strings gives the distance of the $k$-th level-2 block: \begin{align} 
\widetilde{d}_k = d' + \sum_{j=1}^5d_{j,k}. \notag
\end{align} 
We repeat this procedure for all candidate combinations and over all choices of the five level-1 encoded strings and determine the minimum distance $d_k$ among all $\widetilde{d}_k$. For each candidate combination attaining the minimum distance $d_k$, the level-2 encoded string $\mathbf{x}^{(2)}_k$ is calculated from the definitions of $Z^{(2)}_{i',j',k}$ and retained as a level-2 minimum-distance candidate.

The level-3 minimum-distance candidates are obtained similarly. 
However, the number of combinations of  level-2 minimum-distance candidates can become prohibitively large. 
We therefore restrict the number of candidates included in the search. 
For example, we choose candidates for five level-2 encoded strings $\{\mathbf{x}^{(2)}_k\mid k=1,\dots,5\}$ and denote the number of candidates for the $k$-th string by $N_k$. If the total number of candidate combinations $N=N_1N_2N_3N_4N_5$ exceeds $10^5$, we  identify the string with the largest $N_k$, randomly retain one of its candidates, and set the corresponding $N_k$ to one. 
We repeat this reduction until $N\leq10^5$ is reached.
Then, we impose the stabilizer condition  $\mathbf{x}^{(2)}_{6} = -\sum_{k=1}^5\mathbf{x}^{(2)}_{k}$. 
Because the resulting $\mathbf{x}^{(2)}_{6}$ may not be included in the level-2 minimum-distance candidates obtained in the previous step, its distance from the measurement outcomes must be estimated. Rather than searching over all physical codewords corresponding to $\mathbf{x}^{(2)}_{6}$ to determine the global minimum Hamming distance from $\{x_{i,j,6}\mid i,j=1,\dots,6\}$, we adopt a restricted-search procedure. 
Specifically, one of the six level-1 encoded strings constituting $\mathbf{x}^{(2)}_{6}$ is chosen from the minimum-distance candidates obtained during level-1 decoding. 
The remaining five level-1 encoded strings are then determined from the stabilizer constraints and the definitions of the logical operators. The Hamming distance is evaluated for the resulting configuration of six level-1 encoded strings, and the minimum distance is  determined by repeating this procedure over all candidate combinations retained in the restricted search.
To reduce the computational cost, we further limit the number of level-1 candidates considered in the search.
We denote the number of minimum-distance candidates for the $j$-th level-1 encoded string by $M_j$. If $M=M_1+M_2+M_3+M_4+M_5+M_6$ exceeds six, we  identify the string with the largest $M_j$, randomly retain one of its candidates, and set the corresponding $M_j$ to one. 
We repeat this reduction until $M\leq6$ is reached.

Given $\mathbf{x}^{(2)}_{6} = \{x^{(2)}_{i',j',6}|i',j'=1,\dots,4\}$, determined from the stabilizer condition, and $\mathbf{x}^{(1)}_{1,6} = \{x^{(1)}_{i',1,6}|i'=1,\dots,4\}$, chosen from the minimum-distance candidates, we determine the remaining five level-1 encoded strings constituting $\mathbf{x}^{(2)}_{6}$ from the level-1 stabilizer condition and the definitions of the level-2 logical operators as follows: \begin{align} x^{(1)}_{i',2,6} &=x^{(2)}_{i',1,6}-x^{(1)}_{i',1,6}, \notag\\ x^{(1)}_{i',3,6} &=x^{(2)}_{i',2,6}-x^{(1)}_{i',2,6}, \notag\\ x^{(1)}_{i',5,6} &=\sum_{j'}x^{(2)}_{i',j',6}-x^{(1)}_{i',2,6}, \notag\\ x^{(1)}_{i',4,6} &=x^{(2)}_{i',3,6}-x^{(1)}_{i',5,6}, \notag\\ x^{(1)}_{i',6,6} &=x^{(2)}_{i',4,6}-x^{(1)}_{i',5,6}, \end{align} where $i'=1,\dots,4$.
We calculate the minimum Hamming distance $d'_{j,6}$ between the measurement outcomes $\mathbf{x}_{j,6}$ and the physical codewords corresponding to $\mathbf{x}^{(1)}_{j,6}$ for $j=1,\dots,6$.  The total distance for this candidate combination is  $\sum_{j=1}^6d'_{j,6}+\sum_{k=1}^5d_k.$
We repeat the above procedure over the restricted set of level-1 candidate combinations determined by the cutoff on $M$ and retain the minimum distance. We then repeat the entire procedure over the restricted set of level-2 candidate combinations determined by the cutoff on $N$ and determine the overall minimum distance. Finally, we randomly choose one of the candidates attaining the minimum distance and obtain the corresponding level-3 encoded string from the definitions of $Z^{(3)}_{i',j',k'}$.

%

\end{document}